\documentclass{andp2012}
\usepackage[english]{babel}


\title{Different regimes of ultrashort pulse propagation in disordered layered media with resonant loss and gain}

\author[D.\,V. Novitsky]{Denis~V.~Novitsky\inst{1,2,}\footnote{Corresponding author\quad E-mail:~\textsf{dvnovitsky@gmail.com}}}
\author[D. Redka]{Dmitrii~Redka\inst{3}}
\author[A.\,S. Shalin]{Alexander~S.~Shalin\inst{2,4}}
\address[1]{B.~I.~Stepanov Institute of Physics, National
Academy of Sciences of Belarus, 68 Nezavisimosti Avenue, Minsk
220072, Belarus}
\address[2]{ITMO University, 49 Kronverksky Prospekt, St. Petersburg 197101, Russia}
\address[3]{Department of Photonics, Saint-Petersburg State Electrotechnical University, 5 Prof. Popova Street, St. Petersburg 197376, Russia}
\address[4]{Ulyanovsk State University, 42 Lev Tolstoy Street, Ulyanovsk 432017, Russia}

\shortauthors{D.~V.~Novitsky and A.~S.~Shalin} 

\begin{abstract}
Different optical nanostructures containing both loss and gain components attract ever-increasing attention as novel advanced materials and building blocks for a variety of nanophotonic and plasmonic applications. Unique tunable optical signatures of the so-called active metamaterials support their utilizing for sensing, imaging, and signal processing on micro- and nanoscales. However, this tunability requires flexible control over the metamaterials parameters, which could be provided by involving a set of nonlinear interactions. In this paper, we propose a method of governing ultrashort pulses by varying the level of a population difference disorder in a random active metamaterial. This enables to deliver three different interaction regimes: self-induced transparency (low disorder), localization regime (moderate disorder), and light amplification (strong disorder) corresponding to strongly different light pulses speeds. Since this control could be realized via rather plain tools, like simple pump tuning, the proposed disordered medium opens a room of opportunities for designing peculiar active component for a whole set of highly demanded optical applications.
\end{abstract}

\shortabstract 

\begin{document}
\maketitle

\section{Introduction}

Since the first optical metamaterials utilized plasmonic effects and were based on metal-containing nanostructures, the problem of losses has arisen from the very beginning of this field. The idea to add active (gain) components to nanoplasmonic metamaterials to compensate losses was a natural next step \cite{Hess2012}. This concept was successfully applied to negative-refractive-index metamaterials \cite{Wuestner2010,Xiao2010} which were of primary interest in the early investigations in nanophotonics. The topic of active metamaterials has gained great momentum since then. Both a number of active structures considered and a list of effects in these structures increased dramatically. For example, widely utilized hyperbolic metamaterials \cite{Smith2003,Belov2003,Poddubny2013,Shalin2015a,Chebykin2015,Shalin2015b,Ivinskaya2018} supplemented with gain material allow to reach simultaneously spontaneous emission enhancement and outcoupling of high-wavenumber modes \cite{Galfsky2015}, strong coupling \cite{Vaianella2018}, and loss compensation \cite{Novitsky2017}. Other active nanostructured systems include magnetic metamaterials for tunable reflection \cite{Deng2014}, graphene-based random metamaterials \cite{Marini2016}, and hybrid metal-semiconductor metasurfaces for polarization switching and light-beam steering \cite{Cong2018}. There is a multitude of studies on novel laser sources based on active metamaterials, from dark-state lasing in metal-dielectric systems \cite{Droulias2017} to metastructures utilizing superconducting qubits \cite{Asai2015} and randomly dispersed graphene flakes \cite{Marini2016}. The fruitfulness of studies devoted to photonic systems containing both loss and gain is vividly illustrated by the great attention to the so-called PT-symmetric structures in recent years. Although such systems are often developed on the basis of waveguides and cavities (see, e.g., recent reviews \cite{Feng2017,El-Ganainy2018}), there are also proposals of  PT-symmetric metamaterials \cite{Alaeian2014,Novitsky2018b}.

In this paper, we modify the usual loss-gain setting by introducing another variable into system -- disorder. Since the seminal paper by P.~W.~Anderson
\cite{Anderson1958} who predicted the localization of matter waves in disordered lattices, the concept of Anderson localization was transferred to optics and became the starting point of the fruitful field of disordered photonics
\cite{Wiersma2013,Segev2013}. It turned out recently that disordered metamaterials are of great interest not only due to Anderson localization \cite{Gredeskul2012,Scheinfux2017}, but also can serve for observation of topological state transitions \cite{Liu2017}, transmission enhancement \cite{Scheinfux2016}, and wavefront shaping \cite{Jang2018}. 

Although we are concerned here with metamaterial-like structures, it should be mentioned that the disordered photonics with active materials has a long history. The main concern of these studies was perhaps random lasing based on multiple scattering of light in amplifying disordered media. The standard approach to reach random lasing is to use solid-state powders \cite{Noginov2005}, including semiconductors \cite{Cao1999}, and dye solutions with scatterers \cite{Lawandy1994}. More advanced variant is a fiber laser with random distributed feedback \cite{Turitsyn2010,Turitsyn2014,Du2016}. Recently, random lasing in the Anderson localized regime was reported both in 2D disordered photonic crystals \cite{Liu2014} and disordered optical fibers \cite{Abaie2017}. Random lasing systems can be also utilized for switching \cite{Leonetti2013} and nonlinearity-induced modification and competition of laser modes \cite{Liu2003,Stano2013}.

We study the disordered one-dimensional layered structure with resonant loss and gain. Another feature of our consideration is the emphasis on the dynamics of a short pulse propagation. Therefore, our results belong not only to the fields of active and disordered photonics, but also continue the long chain of investigations devoted to the coherent pulse propagation in resonant media \cite{Allen}. In particular, our analysis is naturally connected to the studies of the self-induced transparency (SIT) \cite{McCall1969,Poluektov1975} and related coherent effects such as the formation of population density
gratings \cite{Arkhipov2016,Arkhipov2017} and collisions between solitonic pulses \cite{Afanas'ev1990,Shaw1991,Novitsky2011}.

In the pioneering work by Folli and Conti \cite{Folli2011}, the interplay between SIT and localization was studied for the first time. They considered the disordered multilayer formed by randomly changing layer
thicknesses with incorporated two-level particles. In other words, the disorder was in the background medium of the structure, so that the commonly known localized states of light form and then interact with the SIT pulses supported by the nonlinearity due to the two-level particles. Here we are interested in the systems with disorder in the parameters of the resonant part of the system, whereas the background is uniform. In our recent paper \cite{Novitsky2018}, we have proposed the concept of disordered resonant medium which is a realization of such system. It can be considered as a multilayer metamaterial with the density of active particles randomly varying from layer to layer. We have studied pulse localization in this model and analyzed the transition from SIT to localization as a function of a number of parameters, such as disorder
parameter, mean particle density, total length, and the thickness of constant-density layers. Localization in this framework results in a
strong suppression of the transmission and enhancement of both the
reflection and absorption. In the further study \cite{Novitsky2019}, these observations were used to propose a transmission modulation scheme based on disorder-induced inelasticity of pulse-pulse collisions.

\begin{figure}[t]
\includegraphics[scale=0.5, clip=]{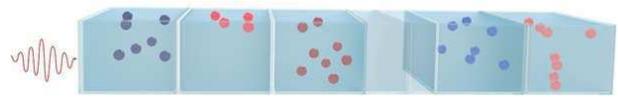}
\caption{\label{fig1} Schematic depiction of the system under consideration. Different shades of blue and red denote different levels of loss and gain, respectively.}
\end{figure}

In contrast to those \textit{passive} systems, here we assume that the density of
active particles is uniform and study another class of disordered resonant media with randomly varying initial population difference. Such a medium can be viewed as a sort of \textit{active} disordered metamaterials and represents a set of layers with the different levels of loss and/or gain. The ordered case is just the uniform resonantly absorbing medium, whereas the introduction of a disorder means appearance of excited layers: the stronger disorder, the larger excitation. Since it is not obvious how such systems response to the external pulsed excitation, we analyze several concrete models of disorder and demonstrate three different regimes of pulse interaction with the system: the SIT regime, the localization regime, and the amplification regime. The first one is observed at a low disorder and is characterized by almost unperturbed slow soliton propagation. The localization regime occurring at an intermediate disorder implies low output and strong absorption of a radiation inside the structure. At a large disorder, the signal rapidly appears back end of the structure, and the total energy of the transmitted and reflected light is amplified. One can switch between these regimes by changing disorder parameter, e.g., via change of pumping. This allows to obtain dramatically different responses (transmission and reflection) at different levels of disorder that can be used as a basis for tunable optical devices.

\section{Governing equations}\label{eqs}

The system considered in this paper consists of a background
dielectric doped with active (two-level) atoms (Fig. \ref{fig1}). Light propagation in
this medium can be described in the semiclassical approximation by
the well-known Maxwell-Bloch equations. Namely, we have differential
equations for the dimensionless electric-field amplitude
$\Omega=(\mu/\hbar \omega) E$ (normalized Rabi frequency), complex
amplitude of the atomic polarization $\rho$, and population
difference between the ground and excited state $w$
\cite{Allen} (see also \cite{Novitsky2011}):
\begin{eqnarray}
\frac{d\rho}{d\tau}&=& i l \Omega w + i \rho \delta - \gamma_2 \rho, \label{dPdtau} \\
\frac{dw}{d\tau}&=&2 i (l^* \Omega^* \rho - \rho^* l \Omega) -
\gamma_1 (w-1),
\label{dNdtau} \\
\frac{\partial^2 \Omega}{\partial \xi^2}&-& n_d^2 \frac{\partial^2
\Omega}{\partial \tau^2}+2 i \frac{\partial \Omega}{\partial \xi}+2
i n_d^2 \frac{\partial \Omega}{\partial
\tau} + (n_d^2-1) \Omega \nonumber \\
&&=3 \epsilon l \left(\frac{\partial^2 \rho}{\partial \tau^2}-2 i
\frac{\partial \rho}{\partial \tau}-\rho\right), \label{Maxdl}
\end{eqnarray}
where $\tau=\omega t$ and $\xi=kz$ are the dimensionless time and
distance, $\mu$ is the dipole moment of the quantum transition,
$\hbar$ is the reduced Planck constant, $\delta=\Delta
\omega/\omega=(\omega_0-\omega)/\omega$ is the normalized frequency
detuning, $\omega$ is the carrier frequency, $\omega_0$ is the
frequency of the quantum transition, $\gamma_{1}=1/(\omega T_{1})$
and $\gamma_{2}=1/(\omega T_{2})$ are the normalized relaxation
rates of population and polarization respectively, and $T_1$ ($T_2$)
is the longitudinal (transverse) relaxation time. The dimensionless
parameter $\epsilon= \omega_L / \omega = 4 \pi \mu^2 C/3 \hbar
\omega$ is responsible for the light-matter coupling, where $C$ is
the density of two-level atoms and $\omega_L$ is the normalized
Lorentz frequency. Quantity $l=(n_d^2+2)/3$ is the local-field
enhancement factor due to the polarization of the background
dielectric with refractive index $n_d$ by the embedded active
particles \cite{Crenshaw2008}. Further we numerically solve Eqs.
(\ref{dPdtau})--(\ref{Maxdl}) using the finite-difference approach
described in Ref. \cite{Novitsky2009}.

We limit our analysis to the one-dimensional system described above
as it is usually done in analysis of SIT and similar effects. The parameters used for calculations are characteristic, e.g., for semiconductor quantum dots as the active particles \cite{Palik,Diels}: the relaxation times $T_1=1$ ns and $T_2=0.1$ ns (much larger than light pulse duration, i.e., we are in the coherent regime), the Lorentz frequency $\omega^0_L = 10^{12}$ s$^{-1}$ (if
not stated otherwise), and the exact resonance ($\delta=0$). The
full thickness of the medium $L=1000 \lambda$. Incident light pulses
have the central wavelength $\lambda=0.8$ $\mu$m and the Gaussian
envelope $\Omega=\Omega_p \exp{(-t^2/2t_p^2)}$ with the pulse
duration of $t_p=50$ fs. For the pulses of such short duration, we
can safely neglect the influence of the near dipole-dipole
interactions between the two-level emitters \cite{Novitsky2010} as
well as the inhomogeneous broadening \cite{Novitsky2014}. The peak
Rabi frequency is given by $\Omega_p=\Omega_0=\lambda/\sqrt{2 \pi} c
t_p$ which corresponds to the pulse area of $2 \pi$ (the $2 \pi$ SIT
pulse). Without loss of generality, we take $n_d=1$; similar results can be obtained for other values of background refractive index as will be shown directly.

The disorder is introduced through the periodical random variations of the initial population difference $w_0=w(t=0)$ along the light propagation direction (see Fig. \ref{fig1}).  Further we consider the concrete model of disorder, which is the simplest example of a general class of disorder allowing to study its influence on the system's response. The model considered implies that the initial population difference in a given layer can take on any possible value. In Supporting Information, our findings are backed with calculations for the two other models of disorder -- the two-valued one with layers either fully inverted or totally unexcited and the generalized one which is a straightforward modification of the previous two. We believe that the behavior of light pulses will be qualitatively the same for different specific models of disorder, so that we can limit ourselves to the simplest ones used in this paper.

\section{Results}\label{mod1}

\begin{figure}[t]
\includegraphics[scale=1., clip=]{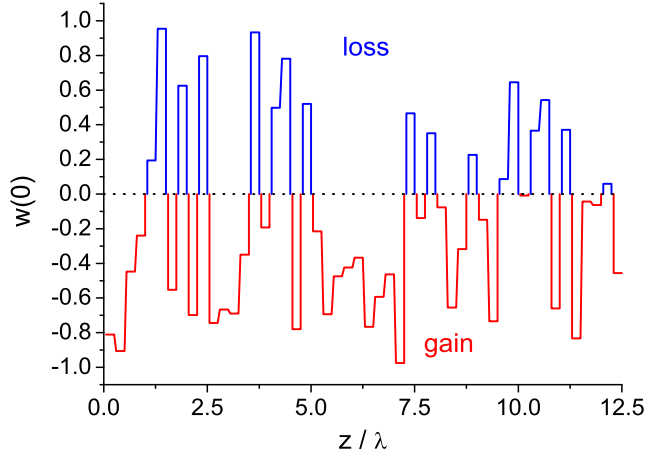}
\caption{\label{fig2} Examples of initial population difference
$w_0$ distributed along the medium for the model of disorder given by Eq. (\ref{randvar1}). The thickness of the layer with constant $w_0$ is $\delta L=\lambda/4$ and the disorder strength is $r=1$.}
\end{figure}

\begin{figure}[t]
\includegraphics[scale=0.95, clip=]{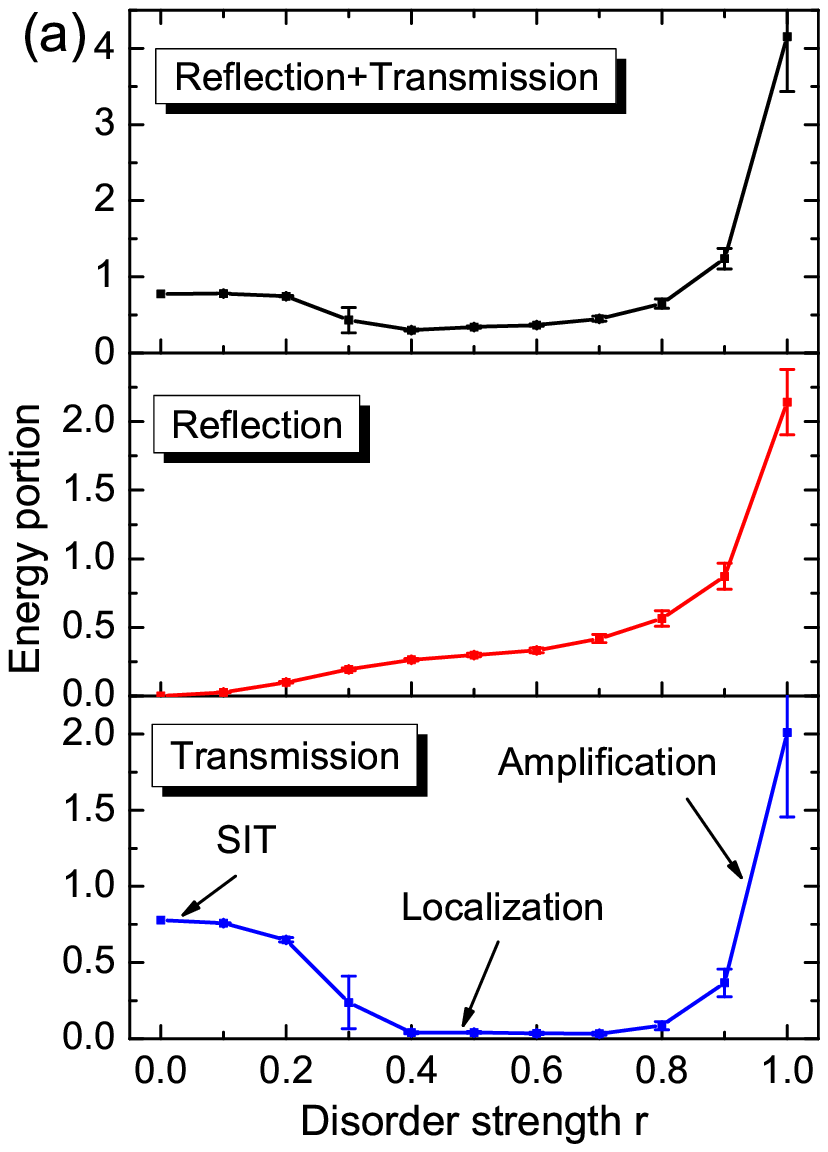}
\includegraphics[scale=0.95, clip=]{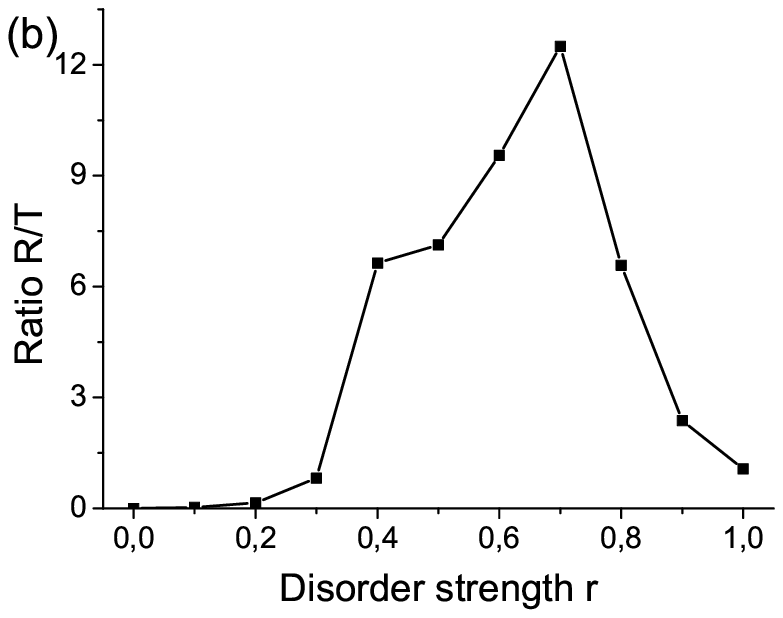}
\caption{\label{fig3} (a) Average output energy of transmitted
(bottom) and reflected (middle) light as well as their sum (top) as
a function of the disorder parameter $r$ calculated within the model of disorder given by Eq. (\ref{randvar1}). Energy averaged over $100$ realizations was calculated for the
time interval $500 t_p$ and was normalized on the input energy. The
layer thickness is $\delta L=\lambda/4$. The error
bars show the unbiased standard deviations for the corresponding
average values. (b) The ratio of average values of reflection and
transmission as a function of $r$.}
\end{figure}

\begin{figure}[t]
\includegraphics[scale=0.8, clip=]{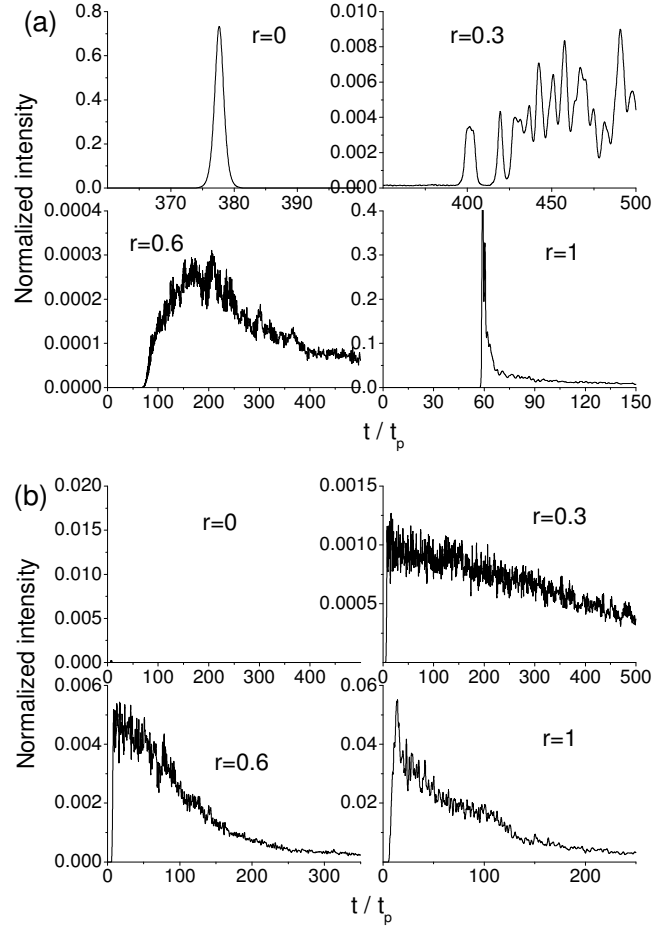} \caption{\label{fig4} Profiles of
(a) transmitted and (b) reflected intensity for different values of
the disorder parameter $r$ calculated within the model of disorder given by Eq. (\ref{randvar1}). Profiles are
averaged over $100$ realizations, the layer thickness
is $\delta L=\lambda/4$.}
\end{figure}

\begin{figure}[t]
\includegraphics[scale=0.95, clip=]{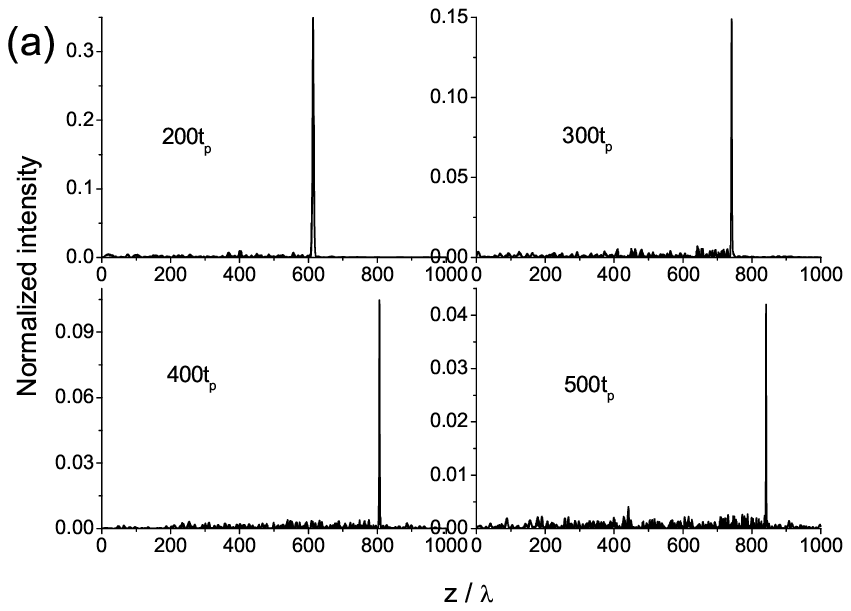}
\includegraphics[scale=0.95, clip=]{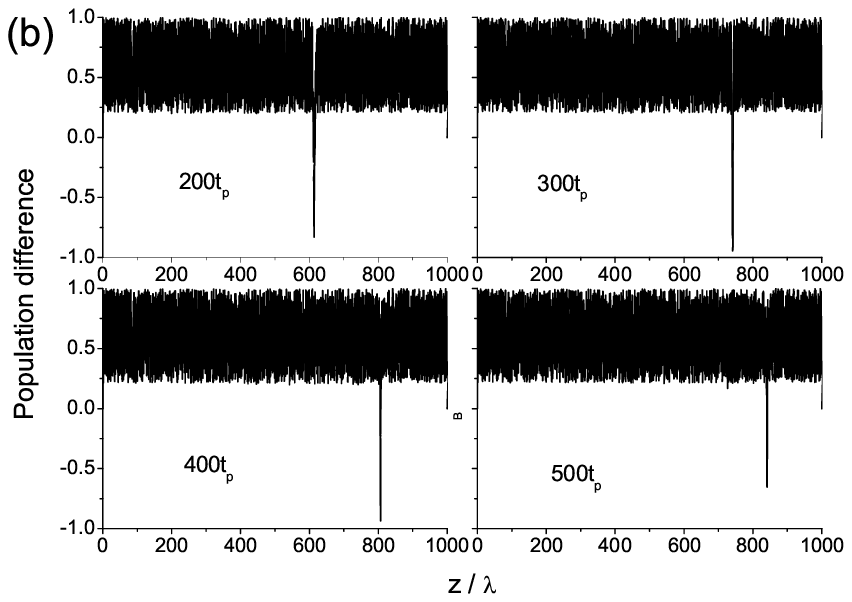}
\caption{\label{fig5} The distributions of (a) light intensity and
(b) population difference at different moments of time for a
realization of disorder with $r=0.4$.}
\end{figure}

\begin{figure}[t]
\includegraphics[scale=1., clip=]{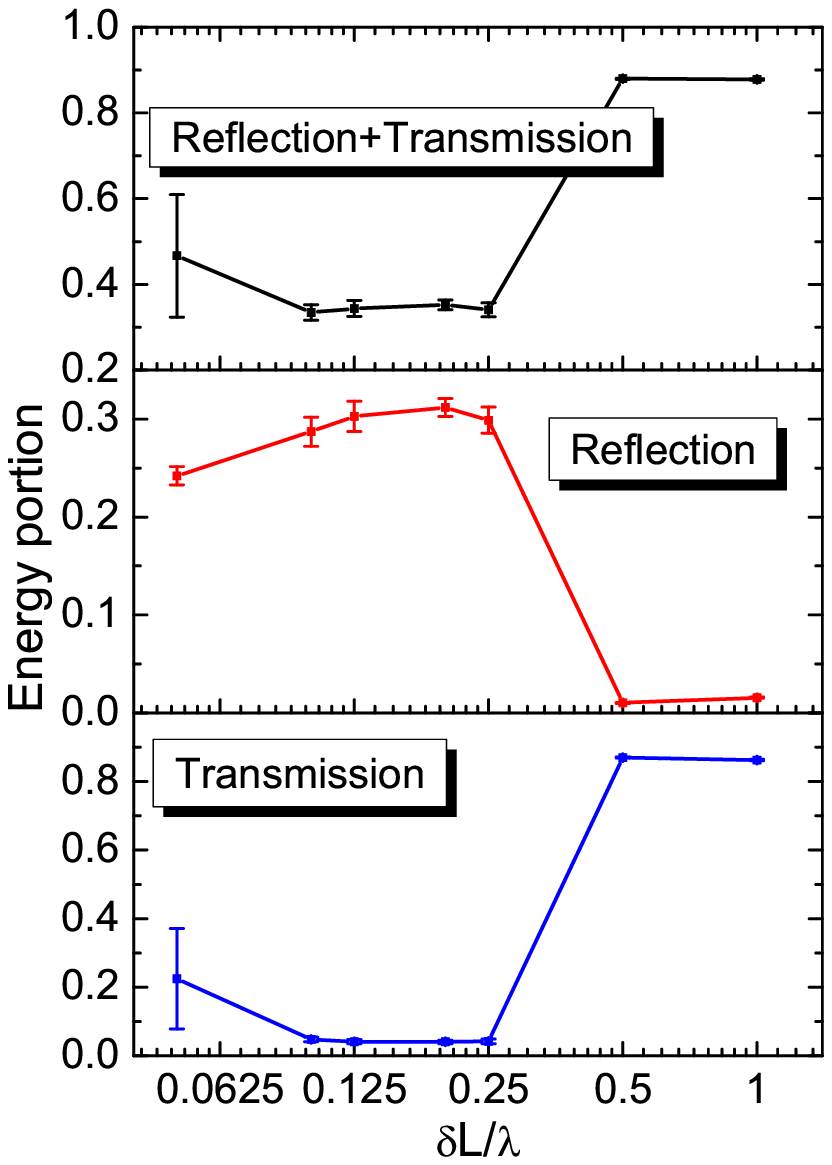} \caption{\label{fig6}
Average output energy of transmitted (bottom) and reflected (middle)
light as well as their sum (top) as a function of the layer thickness $\delta L$. The disorder parameter is $r=0.5$.
Energy averaged over $25$ realizations was calculated for the time
interval $500 t_p$ and was normalized on the input energy.}
\end{figure}

The model of disorder considered here implies that the initial population difference in the $j$th
layer of the medium corresponding to the distance $(j-1) \delta L < z
\leq j \delta L$ is given by
\begin{eqnarray}
w^{(j)}_0 = 1 - 2 \zeta_j r, \label{randvar1}
\end{eqnarray}
where $\zeta_j$ is the random number uniformly distributed in the
range $[0; 1]$ and $r$ is the parameter of the disorder strength. When
$r=0$, we have the trivial case of purely absorbing (lossy) medium
(all $w^{(j)}_0=1$). On the contrary, $r=1$ corresponds to the maximal disorder, when loss and gain have equal probability to appear. In other words, the system can be considered as a multilayer (total thickness
$L$) consisting of slabs (thickness $\delta L$) with different
initial population differences, i.e., different parts of the medium are under different pumping (see Fig. \ref{fig1}). For $r>0.5$, the gain layers with
$w^{(j)}_0<0$ become possible. Thus, the
parameter $r$ not only governs deviation from the ordered case of
pure loss, but also takes on the role of pumping strength resulting
in appearance of gain. An example of distribution governed by Eq.
(\ref{randvar1}) is depicted in Fig. \ref{fig2} for the period of
random density variations $\delta L = \lambda/4$ and maximal
disorder $r=1$.

Figure \ref{fig3}(a) shows the mean values of transmission and
reflection for different levels of disorder calculated in the
framework of the model (\ref{randvar1}). One can distinguish three different regimes of
a pulse interaction with the disordered resonant medium. At low
disorder ($r \leq 0.2$), we have the
\textit{self-induced-transparency (SIT) regime} with high
transmission and the resulting pulse corresponding to a SIT soliton
[see the profile in Fig. \ref{fig4}(a) at $r=0$]. Remind that SIT is the nonlinear effect of high-intensity ultrashort pulse transmission through the resonantly absorbing medium \cite{Allen,McCall1969,Poluektov1975}. The pulse duration should be much shorter than relaxation times of the medium ($t_p << T_1, T_2$), so that the energy absorbed at the front edge of the pulse can be coherently returned to the pulse at its trailing edge. As a result, the solitonic pulse of a specific form can be observed propagating through the medium with low attenuation. This is exactly what we see in Fig. \ref{fig4}(a).

For $r > 0.2$, the
transition to localization occurs with gradual decrease of
transmission and increase of reflection and absorption. In this case, localization means that the radiation gets trapped inside the structure in the form of population inversion \cite{Folli2011,Novitsky2018}. This trapping happens closer to the entrance as disorder grows, so that the slowing down SIT pulses and almost standing population inversion can be observed as shown in Fig. \ref{fig5} for a realization with $r=0.4$. Another sign of transition to localization is the increase of variance \cite{Chabanov2000} -- the statistical value related to the standard deviation shown in figures with the error bars. It is indeed seen that the standard deviation (hence, variance) increases near the transition to the localization regime ($r=0.3$ in Fig. \ref{fig3}(a)), which means strong fluctuations of response from realization to realization.

The average transmission profile at
the point of this transition [Fig. \ref{fig4}(a) at $r=0.3$]
contains several residual peaks due to the presence of strongly
attenuated pulses at the output in some realizations, whereas
the average reflection profile [Fig. \ref{fig4}(b)] has almost linear
slope. For $0.4 \leq r \leq 0.7$, we have the \textit{localization
regime} with very low and almost uniform transmission and reflection
slowly growing with $r$. The average intensity profiles (Fig. \ref{fig4} at
$r=0.6$) have sharp leading edge and exponentially decaying trailing
edge. 

For $r>0.7$, the second transition occurs, from localization
to \textit{amplification regime}, with the rapidly growing transmission
and reflection. The large standard deviations in this regime
indicate that individual realizations of a disorder can strongly deviate from the
mean response, but the amplification is relatively large in all realizations. The average profiles (Fig. \ref{fig4} at $r=1$) have sharp peaks and
flat tails. It is interesting that the transmission peak in this
case appears at the back end much faster than the SIT soliton at $r=0$. The simple
estimation shows that the peak corresponds to the time needed for
photons to almost freely pass through the medium ($\Delta t_{pass} \sim L/c
\sim 50 t_p$), in contrast to SIT regime with slowly moving solitons
($\Delta t_{pass} \sim 380 t_p$). Thus, varying the degree of
disorder, we have the possibility to \textit{control the response
time} of the medium via registering the transmitted signal: from
relatively long times (SIT regime, slow pulse) to infinite time (localization
regime, no pulse) to relatively short times (amplification regime, fast pulse).

Another feature is that changing the disorder parameter (e.g., by
changing pumping) allows us to switch between the regimes with
the prevailing transmission and reflection. Indeed, as we see in Fig.
\ref{fig3}(b), the transition from SIT to localization regime means
that the cumulative reflected energy is much higher than the
transmitted one, the optimal ratio being at $r=0.7$, at the edge of
the localization regime. If we compare not the total energetic
characteristics, but the signal levels shown in Fig. \ref{fig4}, we
can see essentially the same behavior: from the almost zero reflected
signal due to SIT ($r=0$) to the prevailing reflected signal
($r=0.6$) and back to the transmitted signal an order of magnitude stronger than the reflected one ($r=1$). Thus, varying the degree of disorder, we have the
possibility to \textit{control the reflection-to-transmission ratio}
back and forth.

The results reported above were obtained for $\delta L=\lambda/4$ as
the thickness of layers with constant initial population difference. Let
us now discuss the influence of this parameter on the performance of
the disordered medium. We choose the value $r=0.5$ sitting in the middle
of the localization regime and calculate the average transmission and
reflection for different $\delta L$ as shown in Fig. \ref{fig6}.
(Note that we use smaller number of realizations -- $25$ instead of
$100$, -- since our experience evidences that such number is quite
enough to make reliable conclusions). We see that the optimum for
localization (i.e., minimum of total output) is in the range $0.1 \lambda \leq \delta L \leq 0.25
\lambda$ that is in a very good accordance with the results on the problem reported in Ref. \cite{Novitsky2018}.

In Supporting Information, we show that similar results can be obtained for two other models of disorder. In particular, in the framework of the so-called two-valued model, we analyze the importance (and optimization) of the medium thickness and the average concentration of active particles. It is also demonstrated that the same regimes discussed above can be observed in the more realistic structures with non-vacuum host medium.

As to possible experimental implementation, the random distributions of the population difference can be created, e.g., in a side-pumping scheme similar to that utilized in Ref. \cite{Wong2016}. According to our model of disorder, Eq. (4), the random numbers $\zeta_j$ set the random variations of population difference along the radiation propagation direction. This disorder can be introduced into our system within a side-pumping scheme through periodic placing absorbing strips of random thickness on the structure. Different strips will block different portions of pump, so that different regions of the medium will be differently and randomly excited. The disorder parameter $r$, which governs the strength (or degree) of disorder, shows, how large the variations of population difference are. It can be tuned simply by changing pump intensity, so that at the same distribution of random numbers $\zeta_j$, stronger pumping results in
larger amplitude of population-difference variations. Within this conceptual scheme, one can create the distributions of loss and gain with different disorder strengths on demand.

\section{Conclusion}\label{concl}

In conclusion, we have studied the propagation of ultrashort pulses in the
resonant multilayered medium with initial population difference randomly varying along the propagation direction. In contrast to previous considerations of self-induced transparency and localization \cite{Folli2011,Novitsky2018}, we deal with not passive, but active system with disordered loss-gain distribution and uniform background. Calculations performed
for three possible models of the disorder reveal two transitions (and, correspondingly, three different regimes) in such the system
as the disorder parameter grows and the medium starts to strongly
deviate from the uniform absorbing medium. The transitions are: from the self-induced transparency regime to
the localization regime and then to the amplification regime. The
latter appears, only when the population difference can take
negative values and the disorder parameter is large enough. Coexistence of the
three regimes in the same system opens the room of opportunities for
a flexible control over the optical response of the medium. In particular, we show the possibility to govern
the reflection-to-transmission ratio and the speed of a pulse
propagation through the medium by changing the disorder parameter via, e.g., utilizing tunable pumping. Thus, we propose the active multilayered disordered metamaterial-like system (e.g., quantum-dot-based) enabling to switch between different light-interaction regimes, which is of high demand for a variety of applications, such as imaging, signal processing, and as a basis for ultracompact light absorbers (localization regime).

\section*{Supporting Information}
Supporting Information is available from the Wiley Online Library or from the authors.

\begin{acknowledgements}
The work was supported by the Belarusian Republican Foundation for Fundamental Research (Project No. F18-049), the Russian Foundation
for Basic Research (Projects No. 18-02-00414, 18-52-00005, and
18-32-00160), Ministry of Education and Science of
the Russian Federation (GOSZADANIE, Grant No. 3.4982.2017/6.7), and
Government of Russian Federation (Grant No. 08-08). Numerical
simulations of the nonlinear interaction of light with resonant media have been supported by the Russian Science Foundation (Project No. 17-72-10098). The investigation of the profiles of transmitted and reflected intensity for different values of the disorder is partially supported by the Russian Science Foundation
(Project No. 18-72-10127).
\end{acknowledgements}

\end{document}


\title{Supporting Information to ``Different regimes of ultrashort pulse propagation in disordered layered media with resonant loss and gain''}

\author{Denis~V.~Novitsky$^{1,2}$}
\email{dvnovitsky@gmail.com}
\author{Dmitrii~Redka$^3$}
\author{Alexander~S.~Shalin$^{2,4}$}

\affiliation{$^1$B.~I.~Stepanov Institute of Physics, National
Academy of Sciences of Belarus, 68 Nezavisimosti Avenue, Minsk
220072, Belarus \\ $^2$ITMO University, 49 Kronverksky Pr., St. Petersburg
197101, Russia \\ $^3$Department of Photonics, Saint-Petersburg State Electrotechnical University, 5 Prof. Popova Street, St. Petersburg 197376, Russia \\ $^4$Ulyanovsk State University, 42 Lev Tolstoy Street, Ulyanovsk 432017, Russia}

\maketitle

\section{The two-valued model of disorder}\label{mod2}

\begin{figure}[b!]
\includegraphics[scale=1., clip=]{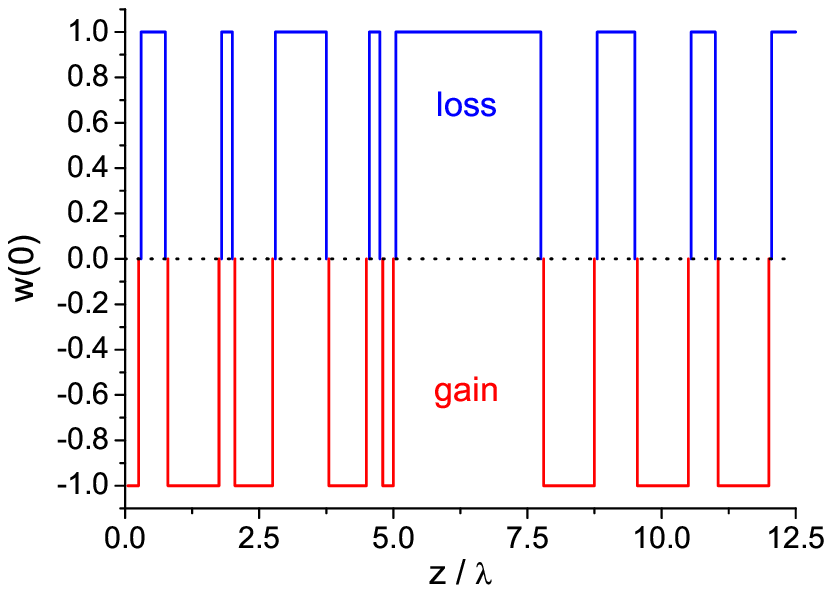}
\caption{\label{figS1} Examples of initial population difference
$w_0$ distributed along the medium for the two-valued model of
disorder given by Eq. (\ref{randvar2}). The thickness of the layer
with constant $w_0$ is $\delta L=\lambda/4$ and the disorder
strength is $r=1$.}
\end{figure}

\begin{figure}[t!]
\includegraphics[scale=0.95, clip=]{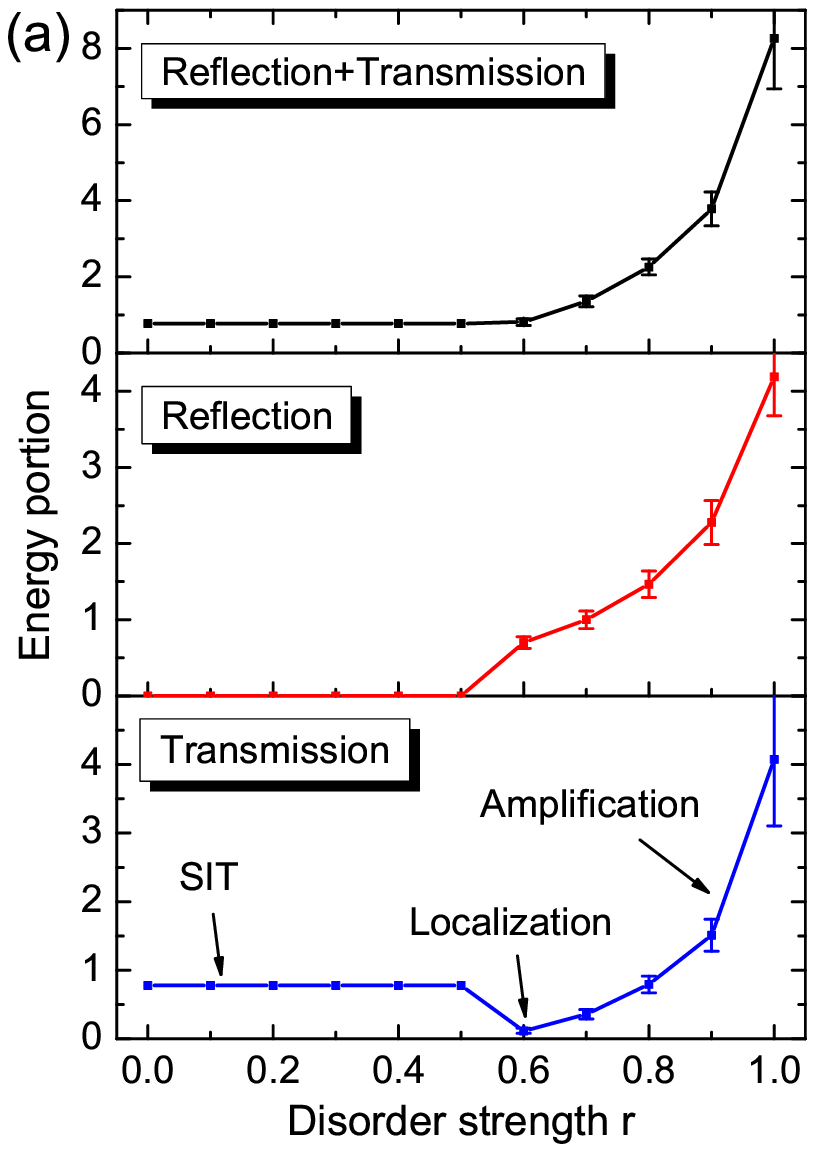}
\includegraphics[scale=0.95, clip=]{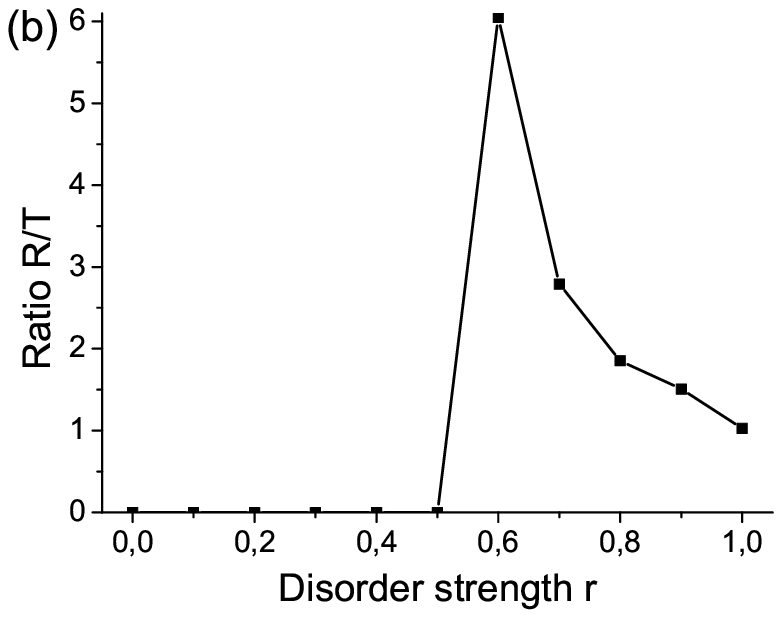}
\caption{\label{figS2} (a) Average output energy of transmitted
(bottom) and reflected (middle) light as well as their sum (top),
depending on the disorder parameter $r$ calculated within the
two-valued model. Energy averaged over $100$ realizations was
calculated for the time interval $500 t_p$ and was normalized on the
input energy. The layer thickness is $\delta L=\lambda/4$. The error
bars show the unbiased standard deviations for the corresponding
average values. (b) The ratio of average values of reflection and
transmission as a function of $r$.}
\end{figure}

\begin{figure}[t!]
\includegraphics[scale=1., clip=]{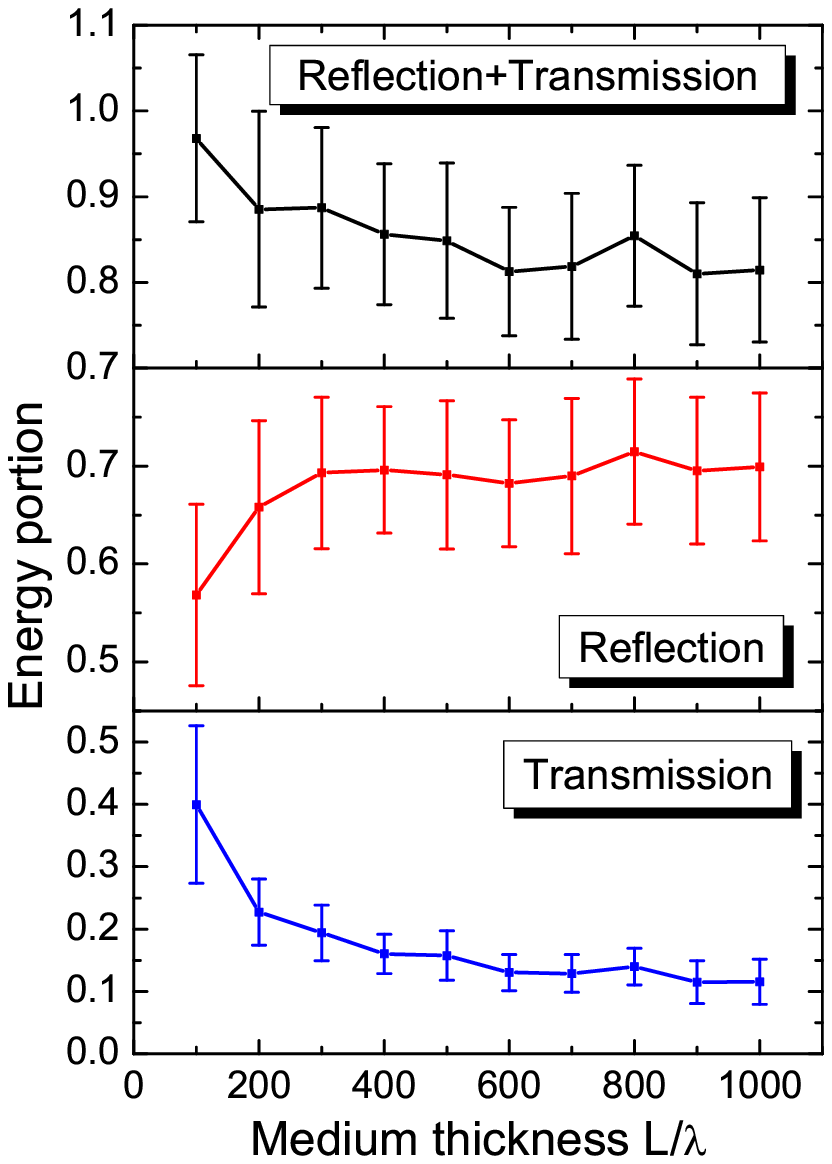} \caption{\label{figS3}
Average output energy of transmitted (bottom) and reflected (middle)
light as well as their sum (top) as a function of the total medium
thickness $L$ calculated within the two-valued model. The disorder
parameter is $r=0.6$. Energy averaged over $25$ realizations was
calculated for the time interval $500 t_p$ and was normalized on the
input energy.}
\end{figure}

\begin{figure*}[t!]
\centering \includegraphics[scale=1.,
clip=]{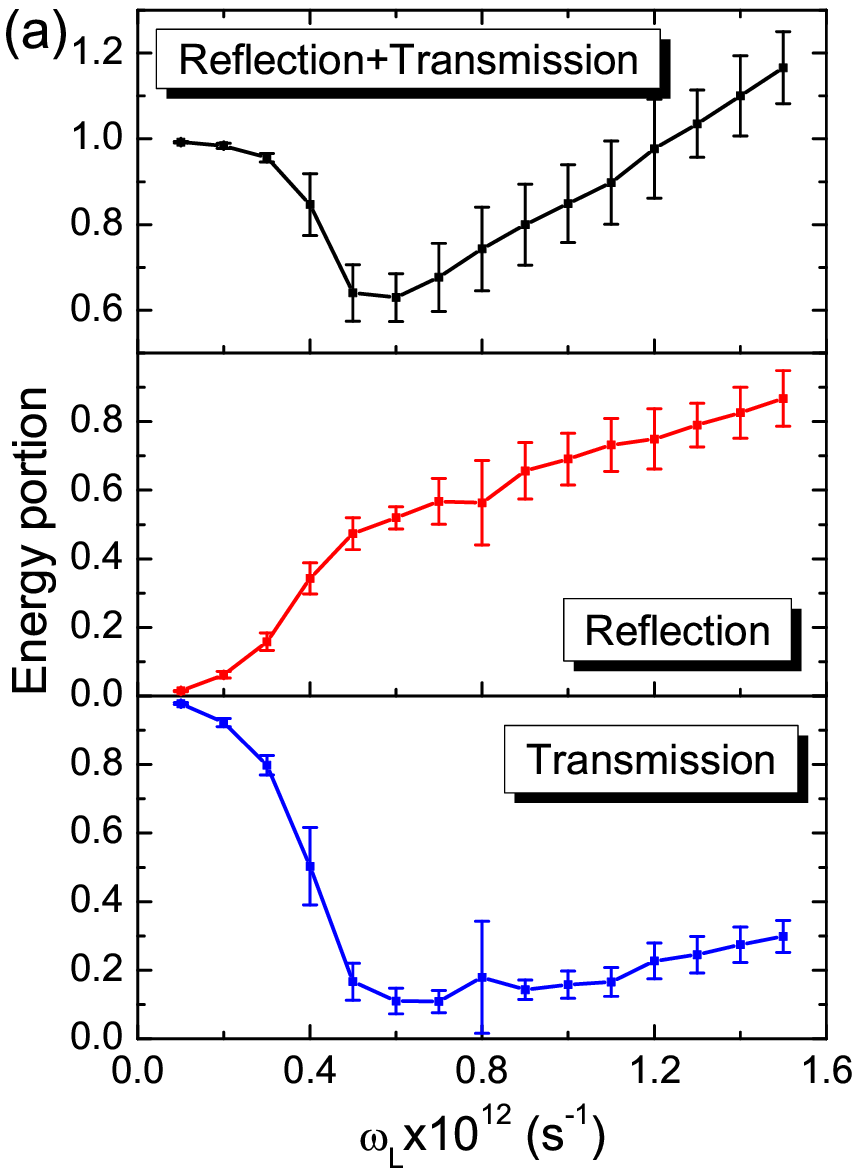}\includegraphics[scale=1., clip=]{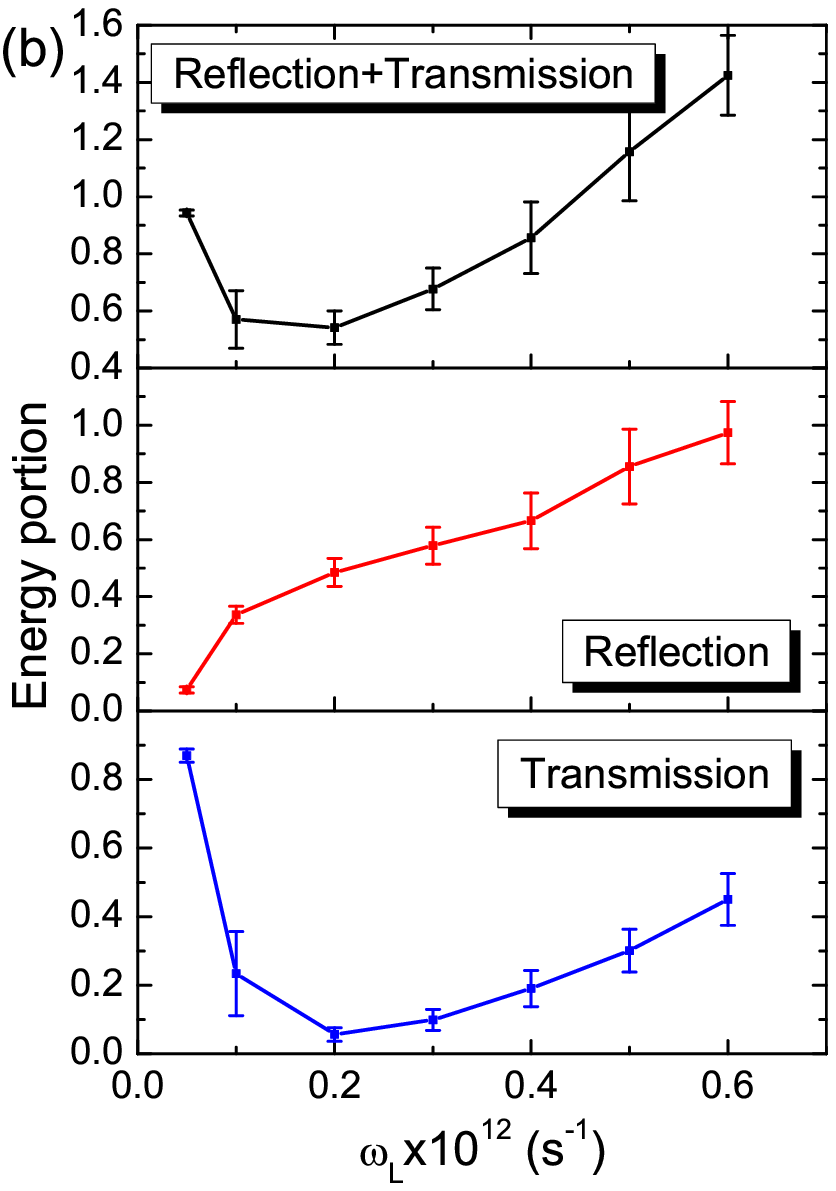}
\caption{\label{figS4} Average output energy of transmitted (bottom)
and reflected (middle) light as well as their sum (top) as a
function of the Lorentz frequency $\omega_L \sim C$ calculated
within the two-valued model. The disorder parameter is $r=0.6$, the
medium thickness is $L=500 \lambda$ the pulse duration is (a)
$t_p=50$ fs, (b) $t_p=200$ fs. Energy averaged over $25$
realizations was calculated for the time interval $500 t_p$ and was
normalized on the input energy.}
\end{figure*}

\begin{figure}[t!]
\includegraphics[scale=1., clip=]{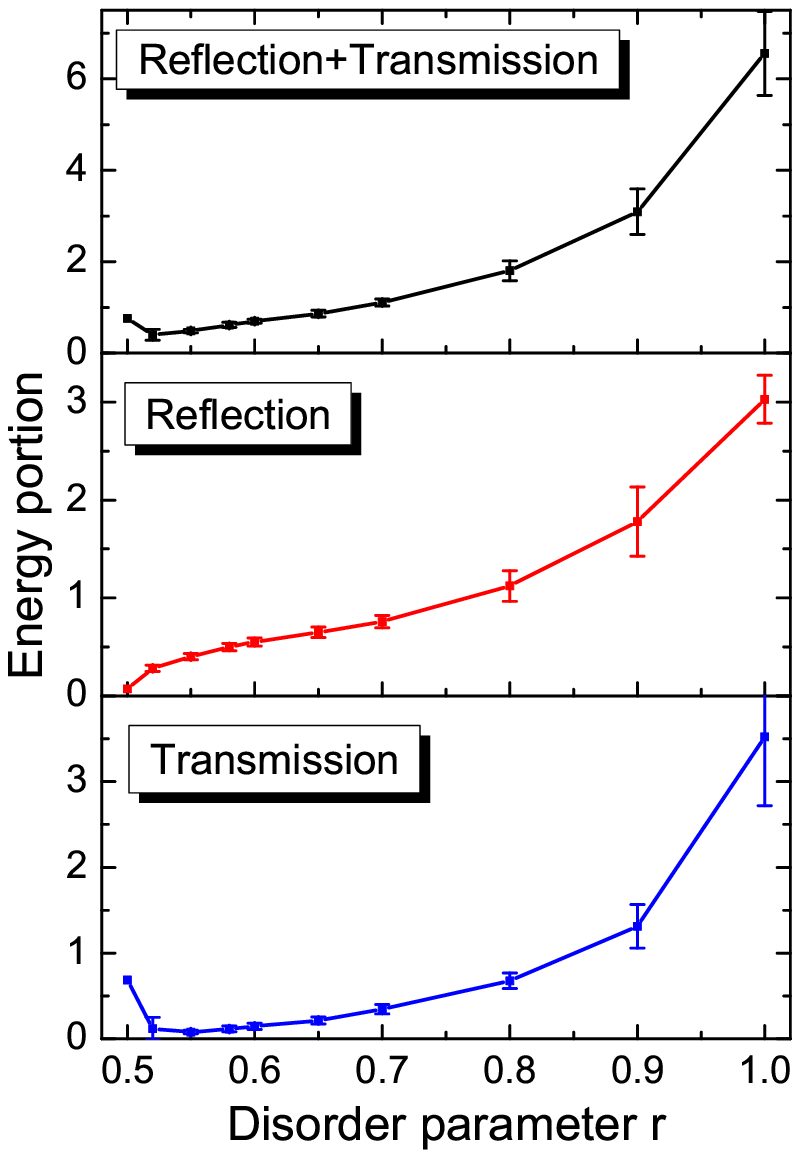}
\caption{\label{figS5} The same as in Fig. \ref{figS2}(a) but for
the background refractive index $n_d=1.5$ and medium thickness
$L=500 \lambda$. Energy averaged over $25$ realizations was
calculated for the time interval $500 t_p$ and was normalized on the
input energy.}
\end{figure}

The two-valued model of disorder can be considered as a
simplification of the one considered in the main text. It is
governed by the initial population difference in the $j$th layer of
the medium corresponding to $(j-1) \delta L < z \leq j \delta L$ as
follows:
\begin{eqnarray}
w^{(j)}_0 = \textrm{sgn}(1 - 2 \zeta_j r), \label{randvar2}
\end{eqnarray}
where $\textrm{sgn}(x)$ is the sign function equal to $+1$ or $-1$
depending on the sign of the argument $x$. In contrast to the model
of the main text, Eq. (\ref{randvar2}) implies that the initial
population difference can have only two values corresponding to loss
($w^{(j)}_0=1$) and gain ($w^{(j)}_0=-1$) layers. As a consequence,
there is no the disorder realizations with only loss layers, since
for $r \leq 0.5$ the system is simply a uniform absorbing medium
with all layers having $w^{(j)}_0=1$, whereas the disorder appears
only for $r>0.5$ and requires the presence of both loss and gain
layers (remind that the disorder parameter characterizes the
probability of the deviation of the medium from the ground state).
An example of the spatial distribution of the initial population
difference obtained in the framework of the two-valued model is
shown in Fig. \ref{figS1}. Such distributions seem to be simpler
than those of Fig. 2 and may be preferable from the experimental
point of view.

Since the disorder manifests itself only for $r>0.5$, it is not
obvious that the localization regime will take place in this model
at all. Figure \ref{figS2}(a) demonstrates that the localization is
indeed possible in the two-valued model at $r=0.6$. For larger
values of the disorder parameter, the system clearly demonstrates
the transition to the amplification regime. Similar to the model of
the main text [compare with Fig. 3(a)], the reflection curve has a
characteristic shape with the slowly changing slope in the
localization regime. Finally, the ratio of the reflection to
transmission [Fig. \ref{figS2}(b)] has also a maximum at $r=0.6$,
i.e., in the localization regime. Thus, one can see that the simple
two-valued model has the same two transitions between three regimes
as the more complex model of the main text, although the
localization regime can be observed in only a very narrow range of
the disorder parameter. This should be taken into account in
experiments with this class of disorder models.

Let us examine the performance of the two-valued model at the other
set of parameters. In Fig. \ref{figS3}, we study the dependence of
the localization ($r=0.6$) on the total thickness of the disordered
medium. We see that the average transmission and reflection remain
almost unchanged for $L \geq 500$. This means that we can take twice
as short medium without destroying the localization effect. Such the
short medium is used to investigate the importance of active
particles density $C$ by varying the Lorentz frequency $\omega_L
\sim C$. The results shown in Fig. \ref{figS4}(a) evidence the
transition to the localization regime (at $r=0.6$) for $\omega_L > 4
\cdot 10^{11}$ s$^{-1}$, so that we can take the density of active
particles diluted in two times in comparison with our initial
consideration. What is also important, for longer (but still
coherent) pulses, the localization threshold substantially decreases
as shown for $t_p=200$ fs in Fig. \ref{figS4}(b). Indeed, we see
localization already for $\omega_L > 1 \cdot 10^{11}$ s$^{-1}$.
Moreover, there is an optimal value of density corresponding to the
minimal transmission, so that the transmission gradually increases
as the density grows.

Finally, in Fig. \ref{figS5}, we show that the same three regimes
can be obtained with non-vacuum host medium. We use for the
calculations a silica-glass-like lossless medium with $n_d=1.5$. We
see the minimal transmission at $r=0.55$ and even a small dip on the
total output curve (the number of $r$ points is higher than in Fig.
\ref{figS2}). Thus, one can obtain similar results in a more
experimentally realistic setup than the system with active particles
in vacuum.

\section{The generalized model of disorder}\label{mod3}

\begin{figure}[t]
\includegraphics[scale=0.95, clip=]{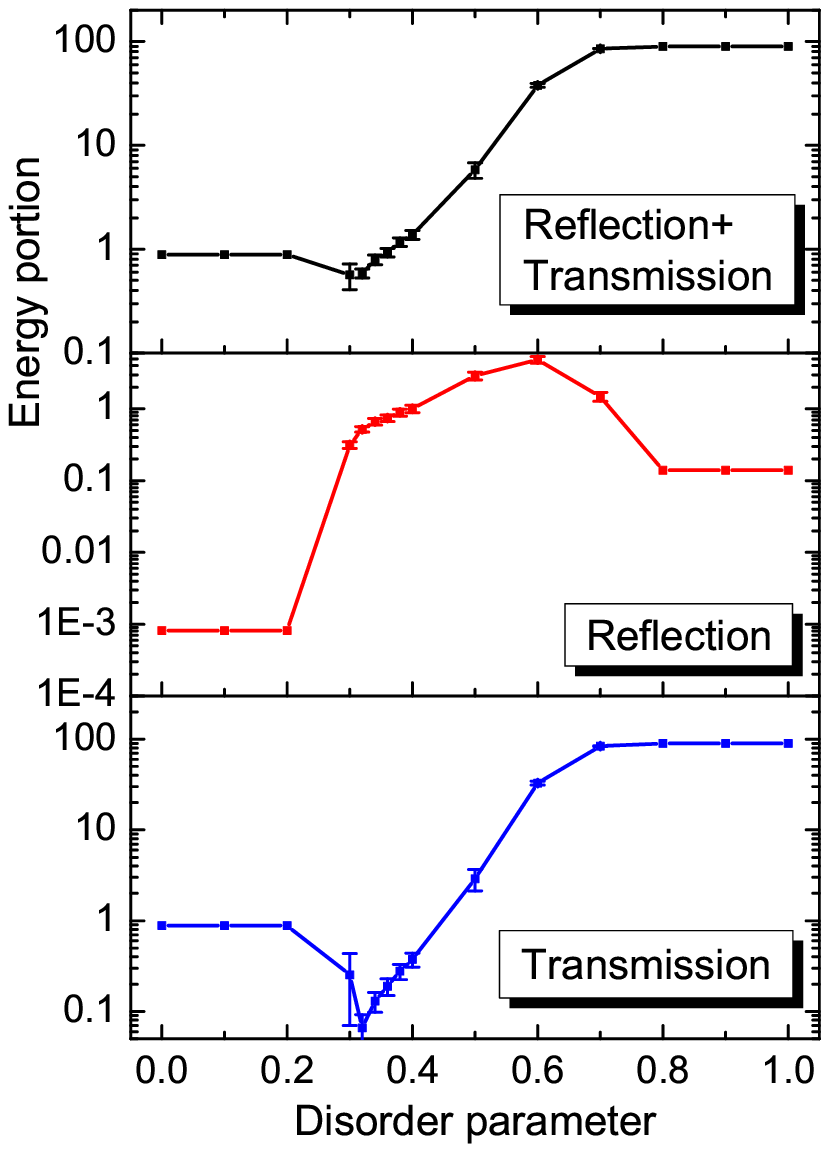}
\caption{\label{figS6} Average output energy of transmitted (bottom)
and reflected (middle) light as well as their sum (top), depending
on the disorder parameter $r$ calculated within the generaluzed
model given by Eq. (\ref{randvar3b}). Energy averaged over $25$
realizations was calculated for the time interval $500 t_p$ and was
normalized on the input energy. The layer thickness is $\delta
L=\lambda/4$, whereas the total medium thickness is $L=500 \lambda$.
The error bars show the unbiased standard deviations for the
corresponding average values. Note the logarithmic scale of the
vertical axis.}
\end{figure}

Two previous models have one common feature: in the limit $r=1$, the
probabilities for the loss or gain layer to appear are equal.
Therefore, it is worth to consider another model of disorder which
we call the generalized model. Within this model, we can transfer
from the entirely loss medium ($r=0$) to the purely gain one
($r=1$). The condition for initial population difference in this
model can be written as
\begin{eqnarray}
w^{(j)}_0 = 1 - 2 r [(2 \zeta_j-1) (r-1)+1], \label{randvar3a}
\end{eqnarray}
or
\begin{eqnarray}
w^{(j)}_0 = \textrm{sgn} \{1 - 2 r [(2 \zeta_j-1) (r-1)+1] \}.
\label{randvar3b}
\end{eqnarray}
These formulas give the $w(0)$ distributions similar to those shown
in Fig. 2 and Fig. \ref{figS1}, respectively. Since in this case we
have quadratic dependence on the disorder parameter, the direct
comparison with the linear laws (4) and (\ref{randvar2}) is not
suitable. However, as the calculations for the case of Eq.
(\ref{randvar3b}) show, the transitions from the SIT to the
localization and then to amplification is observed in this model as
well. A simple estimate shows that in this case, we have the purely
absorbing medium ($w^{(j)}_0=1$) for $r \leq 1-1/\sqrt{2} \approx
0.293$ and the purely amplifying medium ($w^{(j)}_0=-1$) for $r \geq
1/\sqrt{2} \approx 0.707$. The dependencies of the transmitted and
reflected energies on the disorder parameter are demonstrated in
Fig. \ref{figS6}. There is a sharp minimum of transmission (down to
several percent) at $r \approx 0.32$ corroborating the transition to
the localization regime. For a larger $r$, we see almost exponential
growth of the transmission indicating the transition to the
amplification regime. At the same time, the reflection reaches
maximum at $r \approx 0.6$ and then substantially decreases, so that
the purely amplifying medium is mostly transmissive as well as the
purely absorbing one. This can be explained by the fact that in both
these limiting cases, we deal with the ordered medium having much
less reflection (due to SIT effect) than the disordered one by
default.